# Analysis of Ayscough's Telescope Adapted to Use at Sea

Igor Nesterenko

FRIB/NSCL, Michigan State University, East Lansing, MI 48824, USA
Budker Institute of Nuclear Physics, Novosibirsk, 630090, RUSSIA
(corresponding author, e-mail: nesterenko@frib.msu.edu)

**Abstract**
An unambiguous accordance between an objective lens for Ayscough′s telescope ′Adapted to Use at Sea′ and another compound lens which was described by William Eastland in well-known *Quere* during the trial in October of 1765 was revealed. Preliminary analysis of small fragments of *Quere* from Eastland′, Stedman′ and Champneys′ testimonies was carried.

**Introduction**

In modern literature on the history of invention and production of achromatic lenses, James Ayscough is known as optician who gained knowledge about an achromatic compound lenses directly from Chester Moor Hall. It was before the famous Dollond′s experiments with prisms (1757). Also, he was the first who shared these knowledge with subcontractors. For instance in 1752, Ayscough engaged the Clerkenwell optician William Eastland to make several telescopes[1] with a compound lens consisting of flat-convex crown and flat-concave flint glasses. According to Eastland's sworn answer, one of the five defendants,[2] Ayscough wrote out requirements for making compound lenses.

*Quere: Suppose a Crown Glass flat on one side and Convex on the other sufficient to make it 2 foot focus then with a flint Glass flat on one side and Concave on the other sufficient by adding together make it 3 foot.*[3]

At the same time, Ayscough told that the lens had been an invention of Chester Moor Hall Esquire then a counsellor in the Temple and the two of them went to work during 1753 to construct several of the telescopes for themselves and for ′friends′.[4] A few years later, Eastland declared to John Dollond to have made and publicly sold many *such* refracting telescopes.

However, using elementary geometric optics, it can be shown that the flint focal length in the compound lens described in *Quere* should be -6 feet. The ratio of the crown to flint focal lengths is -2/6, or -1/3 instead of about -2/3 − the achromaticity condition established by John Dollond from the prism experiments.[5] In the other words, this configuration is not an achromatic lens for typical crown and flint combinations of the 18th century.[6]

Moreover in 1754, Ayscough published advertisement of *a refracting telescopes, which from several late Tryals at Sea by most experienced and competent Judges, are allowed greatly to excel any other yet made in England.*[7] The detail description of this telescope was published in the same year (Fig.1) and in February issue of the London Magazine[8] on the next year. Both descriptions were identical. Since information about this telescope has been widely disseminated in various sources, including periodicals and even in three languages (English, French and Spanish). Evidently all practicing opticians in London were aware of this Ayscough′ successful project for Royal Navy.

---

[1] B. Gee, ′Francis Watkins and the Dollond Telescope Patent Controversy′, (Ashgate, Farnham, 2014), p.94.

[2] Complaint of Peter Dollond versus Francis Watkins, Addison Smith, William Eastland, Christopher Stedman, James Champneys, 1July – 30 November 1765.

[3] ′The Answer of William Eastland one of the Defendants to Peter Dollond, Complainant′ (29 October 1765). National Archives, Kew, London, Public Record Office MSS, Chancery Proceedings, PRO C12/1956/19.

[4] M. Robischon, ′Scientific Instrument Makers in London During the seventeenth and eighteenth Centuries′, (PhD, University of Michigan, Ann Arbor, 1983), p.289.

[5] D.H. Jaecks, ′An investigation of the eighteenth-century achromatic telescope′, Annals of Science, 67:2, (2010), p.170.

[6] For modern crown BK7 and dense flint SF66 from Schott glass catalogue the achromatic ratio of focal lengths is -0.3254, or about 2% error from -1/3.

[7] J. Ayscough, ′A Short Account of the Eye and Nature of Vision′, 3rd edition, (London, 1754), following p.26.

[8] The London Magazine, or Gentleman's Monthly Intelligencer, V.24, (1755), p.75 or to see Appendix I.

As a rule, many modern scholars tacitly assume that Ayscough′ telescope adapted for use at sea, had a compound and most likely achromatic lens. The fact that this was a compound lens easy to understand from the picture attached to the detail description. But was this lens really achromatic?

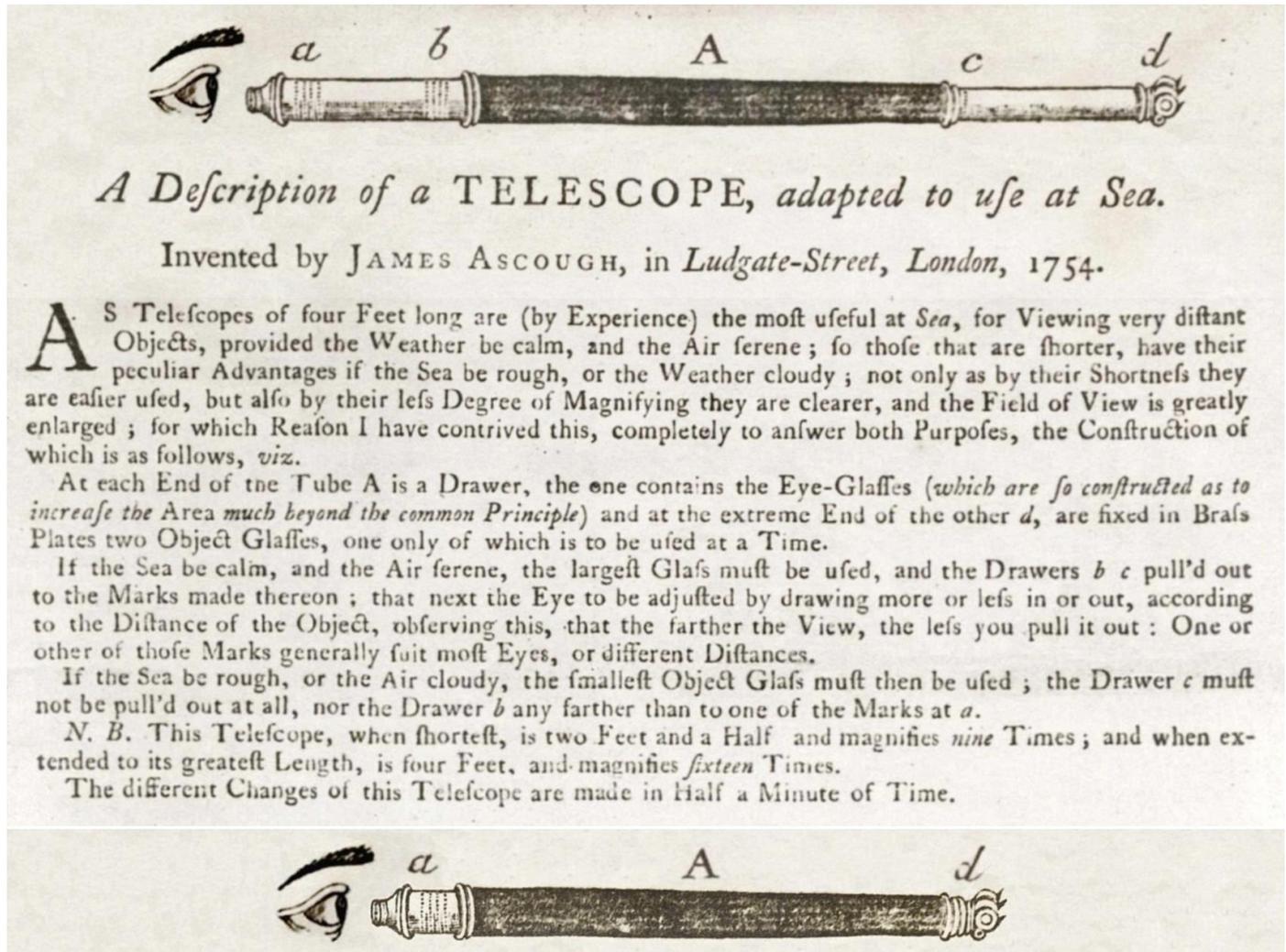

Figure 1. The detail description of telescope which correspond to requirements for using at sea: the top picture of telescope at calm weather; the bottom at windy cloudy weather. The French and Spanish versions of the description have been cut.[9] © Science Museum Group Collection, accession number 1951-687/2.

′Adapted to Use at Sea′ Telescope Analysis

It is known, that two different types of this telescope were made. One with a rotating cubic mount where two singlet objective lenses with short and long focal lengths were installed, and the second with two flip-out lenses (Fig. 2). The first type has a simplest and more robust design. However, this design is outside a scope of this article, as it does not include a compound lens version. The design with two flip-out lenses is more flexible and allows more possible variants including crown-flint compound lenses (Fig. 3). In the version #1 each lenses are used alternately, all the same as in the design with rotating cubic mount. In the version #2 the second lens with short focal length is simply split on two positive lenses. Therefore, a noticeable reduction of chromatic aberration in comparison with version #1 should not be expected. If one of the lenses is made from crown glass and has a short focal length $F_s$, then the additional lens shall have suitable negative focal length to obtain a compound lens with a long focal length $F_l$. Obviously, if the negative lens is made from the same crown glass as the first lens (version #4 and #4a), then a compound lens cannot be achromatic. In this case, achromatism will be achieved when both lenses have equal and opposite focal lengths or theirs ratio is -1.

---

[9] JSTOR link to the full version − https://www.jstor.org/stable/community.26290390

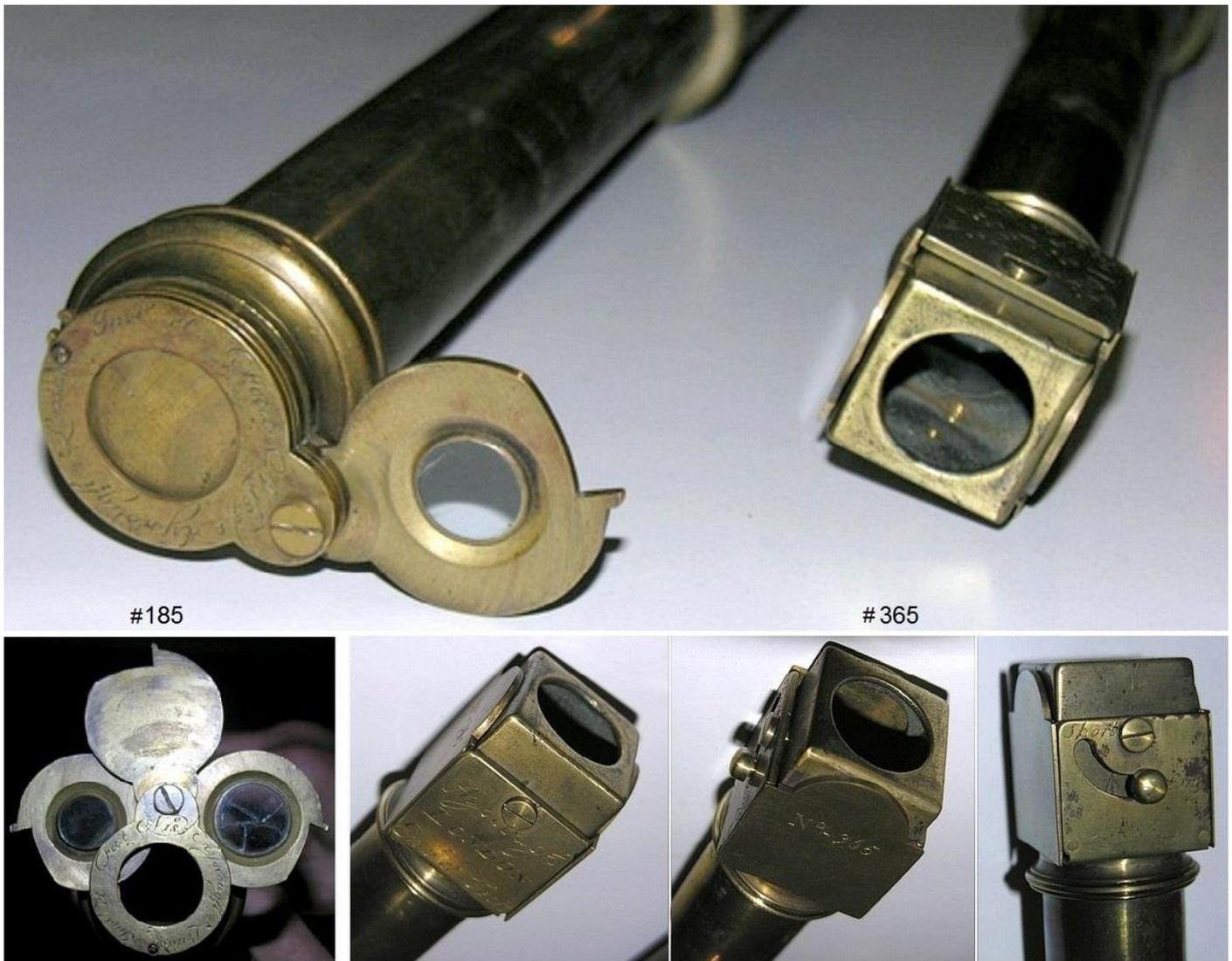

Figure 2. Two different types of the telescope 'Adapted to Use at Sea' with serial numbers 185 (left) and 365 (right), and the same maker signature: 'Ayscough / London / Inv′ et Fecit′, from the collection of Rolf Willach.[10]

Figure 3. Basic crown-forward versions of the design with flip out lenses, and flint-forward versions in Appendix II are shown. The greenish lenses are crown glass, and bluish lenses are flint glass. Here $F_{s,l}$ – short and long focal lengths of the telescope; $F_a$ – focal length of additional lens; $\delta$ – minimal mechanical distance between glasses due to the thickness of flip mounts.

It means that the compound lens will have zero optical power. In general case, when glasses of different types are used, as in versions #5 and #5a, the achromatic condition is expressed by following formula:

$$\frac{F_c}{F_f} = -\frac{v_f}{v_c} \quad (1)$$

Here $v_{c,f}$ – Abbe numbers of crown and flint glasses, and $F_{c,f}$ – theirs focal lengths, respectively.

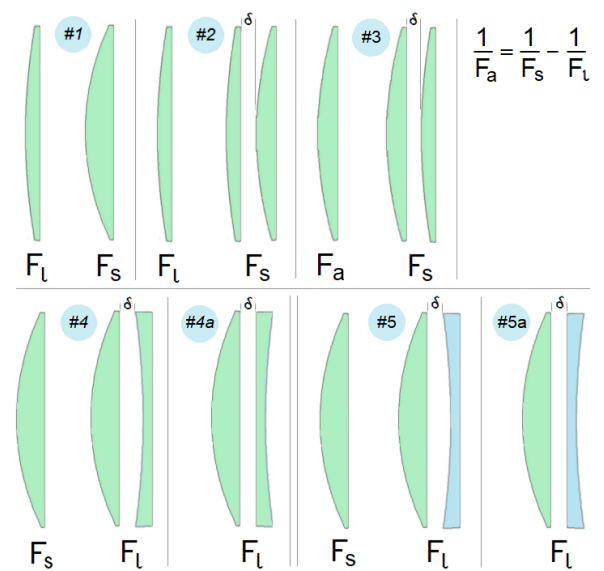

$$\frac{1}{F_a} = \frac{1}{F_s} - \frac{1}{F_l}$$

More detailed analysis of different versions with help of Zemax software can be found in Appendix II.

---

[10] https://dioptrice.com/telescopes/205 (#185) and https://dioptrice.com/telescopes/242 (#365).

The telescope 'Adapted to Use at Sea' consists of two objective lenses with short $F_s$ and long $F_l$ focal lengths and one terrestrial eyepiece with focal length $f_{ep}$. Typically, the eyepiece for terrestrial observations has four glasses: the first two is an erecting system, and the last two is Huygens eyepiece. According to the detail description [N.B.] the telescope magnifications for short and long focal length lenses were $9^x$ ($M_s$) and $16^x$ ($M_l$), respectively. In the other words, the ratio $F_s/F_l = M_s/M_l = 9/16$ is given though the focal lengths themselves are unknown. Based on the design of versions #4th to #6th, it is clear that $F_s = F_c$. The long-focus lens consists of two glasses and its focal length is calculated by the formula[11]: $1/F_l = 1/F_c + 1/F_f$. After multiplying by $F_s$ both sides of this equation and substitute $F_s = F_c$ the final formula for ratio of crown to flint focal lengths was obtained:

$$\frac{F_c}{F_f} = -\frac{M_l - M_s}{M_l} \qquad (2)$$

If we substitute the magnifications in formula (2), then the ratio will be about -0.44, which means chromatic aberration is significantly undercorrected[12], as for typical 18th century crown and flint glasses the achromatic ratio should be approximately -2/3. Thus, we can conclude that the telescope 'Adapted to Use at Sea' was not achromatic even though it could have a crown-flint compound lens.

In addition to magnifications $M_{s,l}$ the detail description contains information about the respective total lengths $T_{s,l}$ of telescope using short-focus or long-focus lenses. The main dimensions of telescope for various conditions are shown on Figure 4. Easy to see from Figures 4B and 4C that the total lengths of telescope with short-focus $T_s = F_s + t_{ep} + 0.5"$ and long-focus $T_l = F_l + t_{ep} + 0.5"$ lenses can be expressed. Subtracting from one equation to another we get:

$$T_l - T_s = F_l - F_s \qquad (3)$$

Based on the fact, that only one eyepiece is used in the telescope, therefore its focal length $f_{ep}$ equal to $F_l/M_l = F_s/M_s$. Substituting this equation into formula (3), we obtain the following formulas for focal lengths:

$$F_s = M_s \cdot \frac{T_l - T_s}{M_l - M_s} \; ; \quad F_l = M_l \cdot \frac{T_l - T_s}{M_l - M_s} \qquad (4)$$

The focal and total lengths of terrestrial eyepiece can be calculated by the formulas:

$$f_{ep} = \frac{T_l - T_s}{M_l - M_s} \; ; \quad t_{ep} = T_l - F_l - 0.5" = T_s - F_s - 0.5" \qquad (5)$$

Substituting in formulas (4) the total lengths $T_s = 2.5'$; $T_l = 4.0'$ and respective to them magnifications ($9^x$ and $16^x$), we get the following values for short and long focal lengths (in feet): $F_s = 1.9_3'$; $F_l = 3.4_3'$. From formulas (5) the focal and total lengths of terrestrial eyepiece (in inches) are $2.5_7"$ and $6.3_6"$ respectively[13]. The values obtained for the focal lengths of both objective lenses are close to those given in *Quere*, especially for the short-focus crown lens. Moreover, the focal length of long-focus compound lens is quite consistent with the entry '3.+' in the sworn answer of James Champneys, and after rounding to nearest integer will also correspond to the value in testimonies of other defendants (Fig. 5). Note that all formulas above based on the thin-lens approximation. Therefore, it should be expected that simulation results of a real lenses with non-zero thickness and distance between components will be slightly different.

---

[11] For versions #4 and #4a it is better to use the following designations $F_{1c}$ and $F_{2c}$ – the focal lengths of first and second crown lenses instead of $F_c$ and $F_f$ respectively.

[12] The achromatic ratio for modern crown / flint pairs: N-BK7 / SF3 and K7 / N-SF14 is about -0.439. Refractive indices for d-line (0.5876 μm) of flint glasses SF3 and N-SF14 are 1.7400 and 1.7618 respectively. These values were unattainable in 18th century.

[13] The last digits in calculated values are small due to unreliability associated with a measurement accuracy and subsequent rounding of the initial data. More information on possible tolerances of calculated values can be found in Appendix I.

Figure 4. A – The telescope 'Adapted to Use at Sea' is retracted; B – it is focused on image of short-focus lens; C – it is focused on image of long-focus lens.
IP is position of Image Plane and EP is Eye Pupil position.
Here $L_0$ – main tube length; $X$ – drawtube stroke length at objective lens end; $t_{ep}$ – total length of terrestrial eyepiece, equal to the distance between image plane (IP) and eye pupil (EP); $F_{s,l}$ – short and long focal lengths respectively; $T_{s,l}$ – total lengths of telescope for short-focus and long-focus lenses; $X_{s,l}$ – positions of image plane relative to the eye end of main tube; $M_{s,l}$ – telescope magnifications; $t_{o,e}$ – lengths of draw-tube holders at objective lens and eye ends of main tube, respectively.

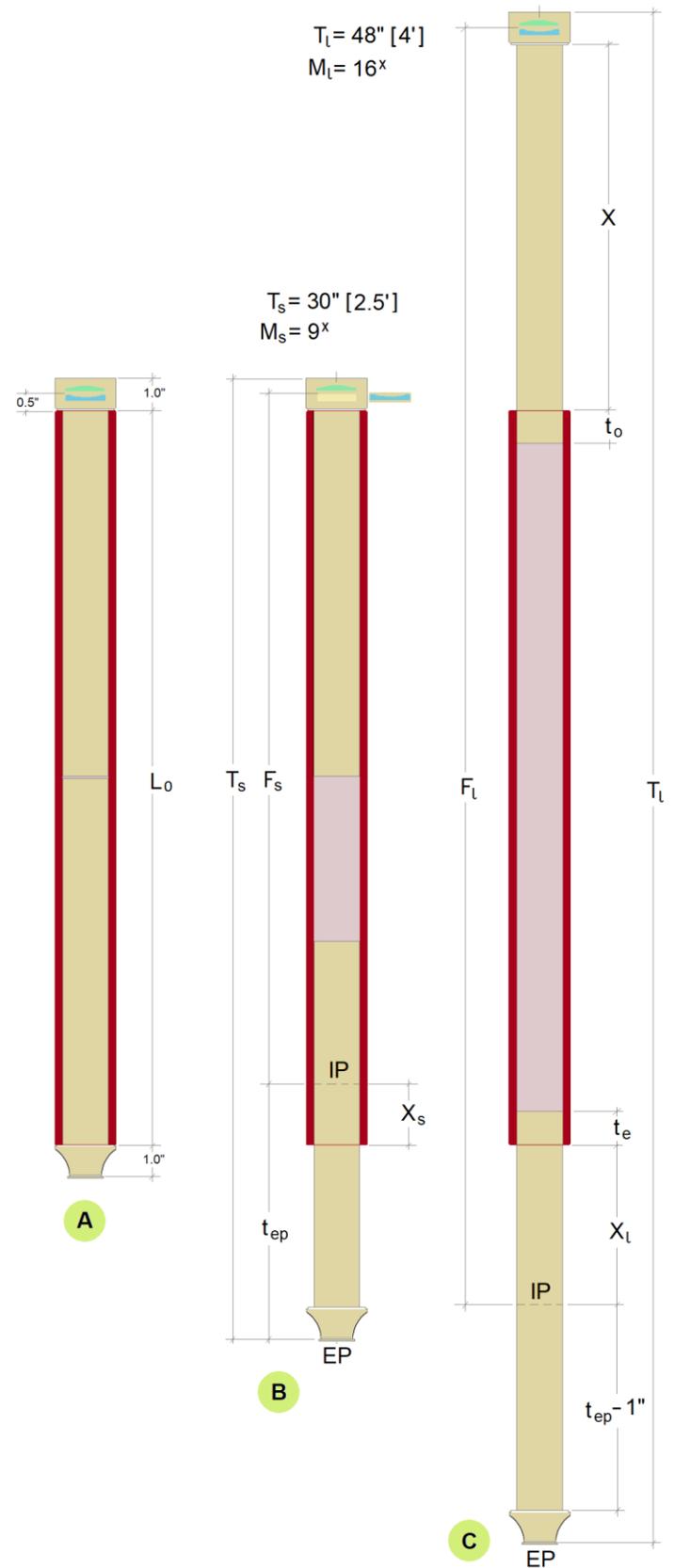

The length of main tube[14] is:

$$L_0 = (T_l - 2" + t_o + t_e)/2$$

With such length of main tube, the long-focus lens can only be focused on distant objects. For focusing on close objects the stroke length of draw-tube from eye side should be increased simultaneously with the main tube length. Accordingly, in the formulas below, $L_0$ should be replaced by more appropriate length.
The image plane position relative to the eye-end of main tube for short-focus lens is:

$$X_s = L_0 - F_s + 0.5"$$

For case with long-focus lens the image plane position $X_l$ can be expressed as a sum with the stroke length of draw-tube $X$ from lens side:

$$X + X_l = F_l - L_0 - 0.5"$$

In the other words, $X$ and $X_l$ values can be freely redistributed within the calculated amount.

**Quere**
The pictures of part original testimonies of Eastland, Stedman and Champneys were presented in work.[15] It turned out that some of the original numbers in *Quere* had been scratched out and altered (Fig. 5). The room in text of Eastland' testimony under '2' is enough for only one digit. Obviously, it is not 0 or 3, and also not 2. It will be strange scratched out 2 and after replace on 2 again. Consequently, it could be only '1', before it was replaced.

---

[14] If the formula contains a number with the symbol ' or ", then all dimensions should be expressed in feet or inches, respectively.
[15] H. Zuidervaart, T. Cocquyt, 'The Early Development of the Achromatic Telescope Revisited', Nuncius, 33, (2018).

Figure 5. Preliminary analysis of Figs. 3A, 3B and 3C.[16] Colored lines indicate spaces between words. Lines of the same color have the same length. Translucent red symbols indicate possible original numbers which were altered, and '$x$' means an arbitrary digit.

If so, then the focal length of flint lens should be -1.5 feet to get 3 feet for a compound lens. In this case, the ratio of crown to flint focal lengths is -2/3, which satisfies the required achromatic condition. However, these values did not correspond to reality – to telescopes 'Adapted to Use at Sea' which were made and publicly sold before 1758.

In texts of Stedman' and Champneys' testimonies the original values under '2.' were also scratched out. In this case, it is impossible to predict what the numbers were previously. Base on visible free space, it can be argued, they looked like '$x.x$', where '$x$' means arbitrary digit.[17] The focal length value of compound lens in Champneys' text was also altered. Most likely, only the fractional part was replaced by '+'. This means that from Champneys' point of view the focal length should be a little longer than 3 feet.

By all appearances, the above described changes in the three testimonies were caused by the defendant's attempts to present a telescope adapted for use at sea as an achromatic with crown-flint compound lens. As a result of litigation, the actual telescope parameters were revealed and included in the final court documents. In the other words, changes in *Quere* text is not the result of 'clerk error', most likely, this is a critical mistake of the defendants themselves.

In order to confirm or refute this hypothesis, it is necessary to conduct a handwriting expertise in these sworn answers. If this investigation reveals a handwriting at least one of the defendants (especially Eastland) then this will unambiguously confirm the hypothesis – there was an attempt on the part of defendants to mislead the court, which was eventually terminated and now we see traces of this event in the authentic documents. In addition to handwriting expertise, order of consideration of these sworn answers in court is also important.

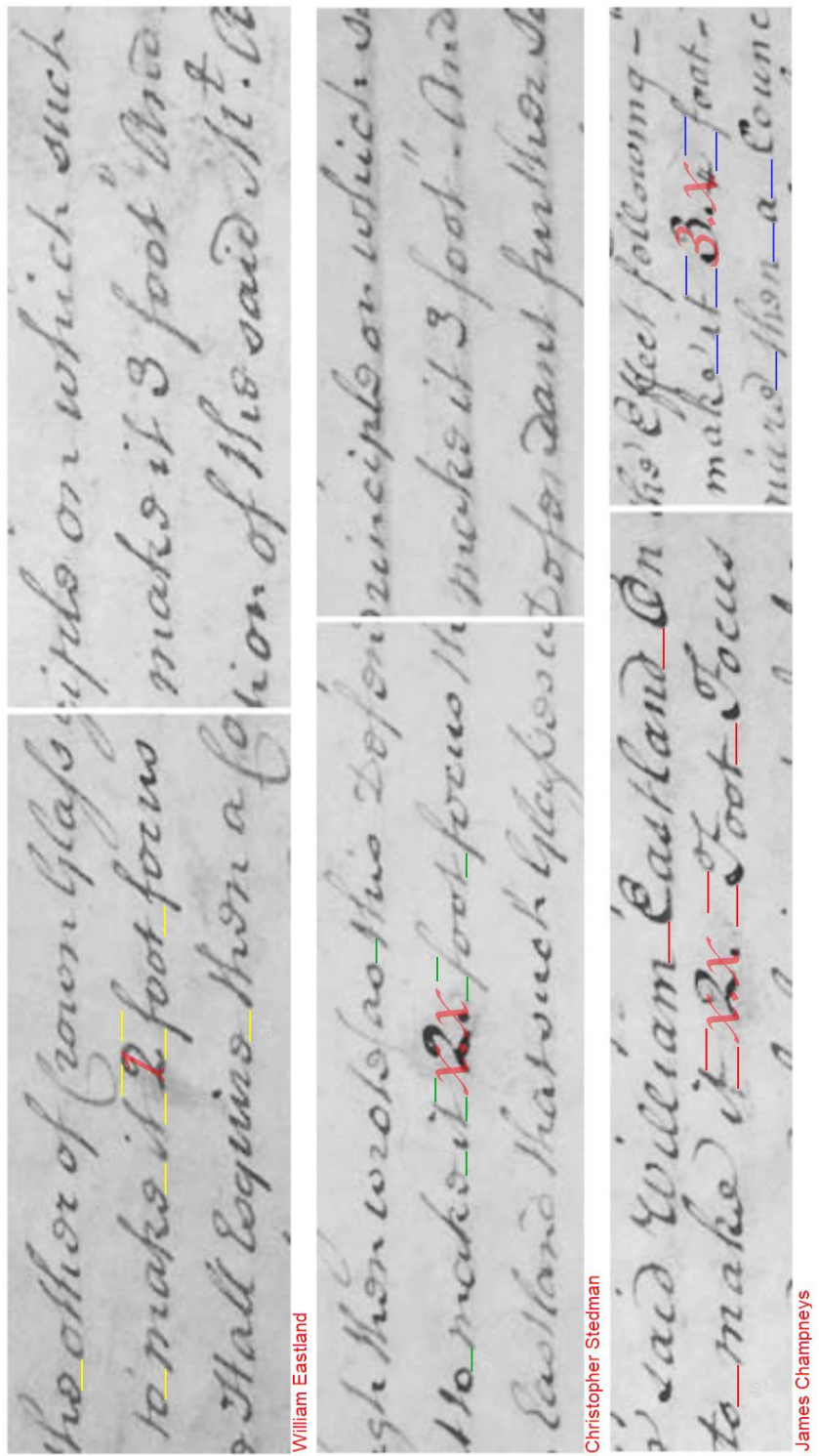

---

[16] Ibid. (reference 15), pp. 286-287.
[17] Of course, the integer part of number '$x.x$' could be 1 or 2, most likely 1, more details in Appendix I.

**Summary**

An analysis of the telescope ′Adapted to Use at Sea′ was carried out. Various possible versions of such telescope was considered. It was shown that the telescope could not be achromatic even with a long-focus compound lens for usual crown and flint glasses of the 18$^{th}$ century.

The calculated focal lengths for both telescope objectives correspond to the rounded values given in *Quere*, which was repeated three time in the sworn answers of defendants – Eastland, Stedman and Champneys. Undoubtedly, it is required to measure the main parameters of surviving telescopes for sea, especially those that have initial serial numbers.

In connection with the recently revealed erasures in the authentic court documents, a preliminary analysis of small original fragments from *Quere* was carried out and a hypothesis for their appearance was proposed. Obviously, further analysis of the full original texts and their handwriting expertise are required.

Despite the fact that Ayscough′ telescope for sea was removed from the list of early achromats, the question about availability of commercial achromats prior to 1758 is still open. Here, it should be recalled his other telescope which *was about 18 inches long and would shew Jupiter's Satellites clearly & distinctly*. Ramsden mentioned this in 1789 during a discussion about the invention of achromatic telescope in the Royal Society. This telescope is significantly different in terms of lens parameters from the telescope for sea. Apparently, it was made in a single copy and was in Ayscough' possession, because in numerous lawsuits no one mentioned publicly sold such refracting telescopes.

**References**

1. B. Gee, ′*Francis Watkins and the Dollond Telescope Patent Controversy*′, (Ashgate, Farnham, 2014)
2. ′*The Answer of William Eastland one of the Defendants to Peter Dollond, Complainant*′, 29 October 1765, National Archives, Kew, London, Public Record Office MSS, Chancery Proceedings, PRO C12/1956/19.
3. M. Robischon, ′*Scientific Instrument Makers in London During the seventeenth and eighteenth Centuries*′, (PhD, University of Michigan, Ann Arbor, 1983).
4. D.H. Jaecks, ′*An investigation of the eighteenth-century achromatic telescope*′, Annals of Science, 67:2, (2010).
5. J. Ayscough, ′*A Short Account of the Eye and Nature of Vision*′, 3rd edition, (London, 1754).
6. The London Magazine, or Gentleman's Monthly Intelligencer, V.24, (1755).
7. H. Zuidervaart, T. Cocquyt, ′*The Early Development of the Achromatic Telescope Revisited*′, Nuncius, 33, (2018).
8. D.D. Maksutov, ′*Astronomical Optics*′, 2nd edition, (Nauka, Leningrad, 1979).

**Appendix I**

According to the telescope description (Fig. 1 or Fig. I-1) its total length is two and a half feet when using a short-focus lens. So, we have to assume that all lengths have been rounded to the nearest half foot. It means that possible length range for a short-focus lens is $2.26' \leq T_s \leq 2.74'$ or $T_s \in [2.26', 2.74']$, and for a long-focus lens is $T_l \in [3.76', 4.24']$. Their corresponding magnifications have been rounded to the integer values $9^x$ and $16^x$. Therefore, the magnification ranges should be chosen in accordance with the mathematical rules for rounding to integer numbers. That means, magnification range with a short-focus lens is $M_s \in [8.52^x, 9.48^x]$[18] and with a long-focus lens is $M_l \in [15.52^x, 16.48^x]$. It is obvious that the ranges significantly exceed a possible measurement errors, so they are not taken in to account. Or it can be supposed that measurement errors are already included in these ranges. Also, the selected ranges for total lengths cover requirements for different viewing distances and are suitable for most eyes.

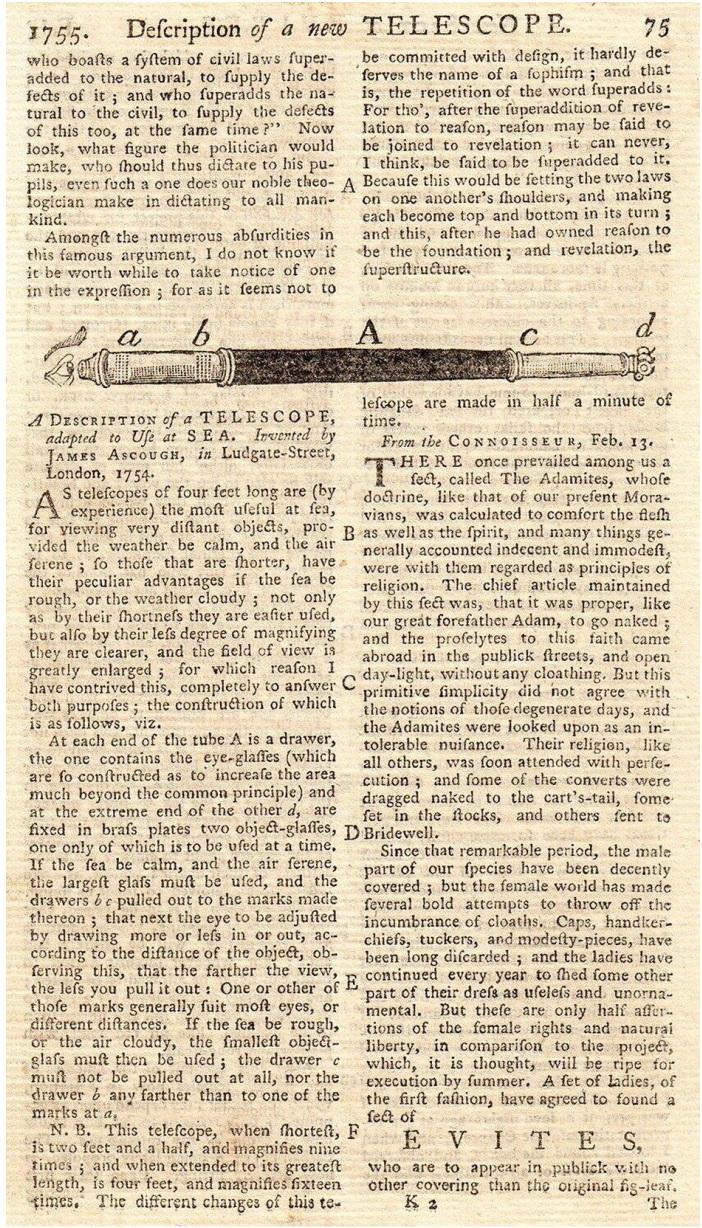

Figure I-1. Image of page 75 from 'The London Magazine' (February, 1755) with description of a new Sea Telescope.[19]

Using formulas (4) together with the declared above ranges for total lengths $T_{s,l}$ and corresponding to them magnifications $M_{s,l}$, we can calculate ranges of short and long focal lengths. All calculated values are shown by red or blue dots on the plots (Fig. I-2). Each dot is one of possible variants for the telescope. If we choose steps by the ranges equal to 0.02 for total length, and for magnification equal to 0.03, then the total number of variants will be 680625. They cover the area from 1.1′ to 3.1′ at short focal length and from 2.1′ to 5.1′ at long focal length. This is a quite representative set for both statistical and dimensional calculations. However, it should be taken into account that the selected ranges of initial parameters are very wide. So, some combinations give unacceptable cases, when a focal length of Huygens eyepiece is more than a total length of terrestrial eyepiece sometimes having a negative length. Obvious, these combinations should be excluded from consideration as unrealistic. They are shown by blue dots (Fig. I-2A). Just as obvious, the sum of a terrestrial eyepiece length and an objective lens focal length is approximately equal to the total length of telescope. Any terrestrial eyepiece consists of an erecting system and, as rule, a Huygens eyepiece. Both parts have approximately the same length. The Huygens eyepiece has a mechanical length greater than its focal length. Therefore, the total length of terrestrial eyepiece can't be shorter than two focal length. Usually, its total length is about three times the focal length. To avoid losing telescope variants, the length of terrestrial eyepiece should be taken 2.5 times its focal length or a little more. Using this selection condition for both objective lenses, number of variants is decreased to 357813. These variants are shown on the plots by red dots.[20] Any variants marked by red dots should be considered equally probable.

---

[18] The range is a little narrower than according to the rounding rules due to desire for symmetry with respect to the declared value, and on the other hand, existing limitations of Mathcad software.

[19] https://babel.hathitrust.org/cgi/pt?id=mdp.39015021267748&view=1up&seq=77&skin=2021

Nowadays we know that the original focal lengths in *Quere* were altered on '2′' or '2.′', and also on '3.+′' in Champneys′ testimony (Fig. 5). Naturally to assume that these values were rounded after removing original numbers. Therefore, according to *Quere* description the range for short-focus lens should be $F_s \in [1.51', 2.49']$, and for long-focus lens is $F_l \in [2.51', 3.49']$ – the bluish areas on the plots. The number of red dots in the intersection area is 279880 or more than 78% of the full red dot set. The median values for this red dot subset are $1.7_5'$ for short and $3.1_4'$ for long focal length. The average and standard deviation are $1.7_5' \pm 0.1_3'$ and $3.1_2' \pm 0.2_1'$, respectively.

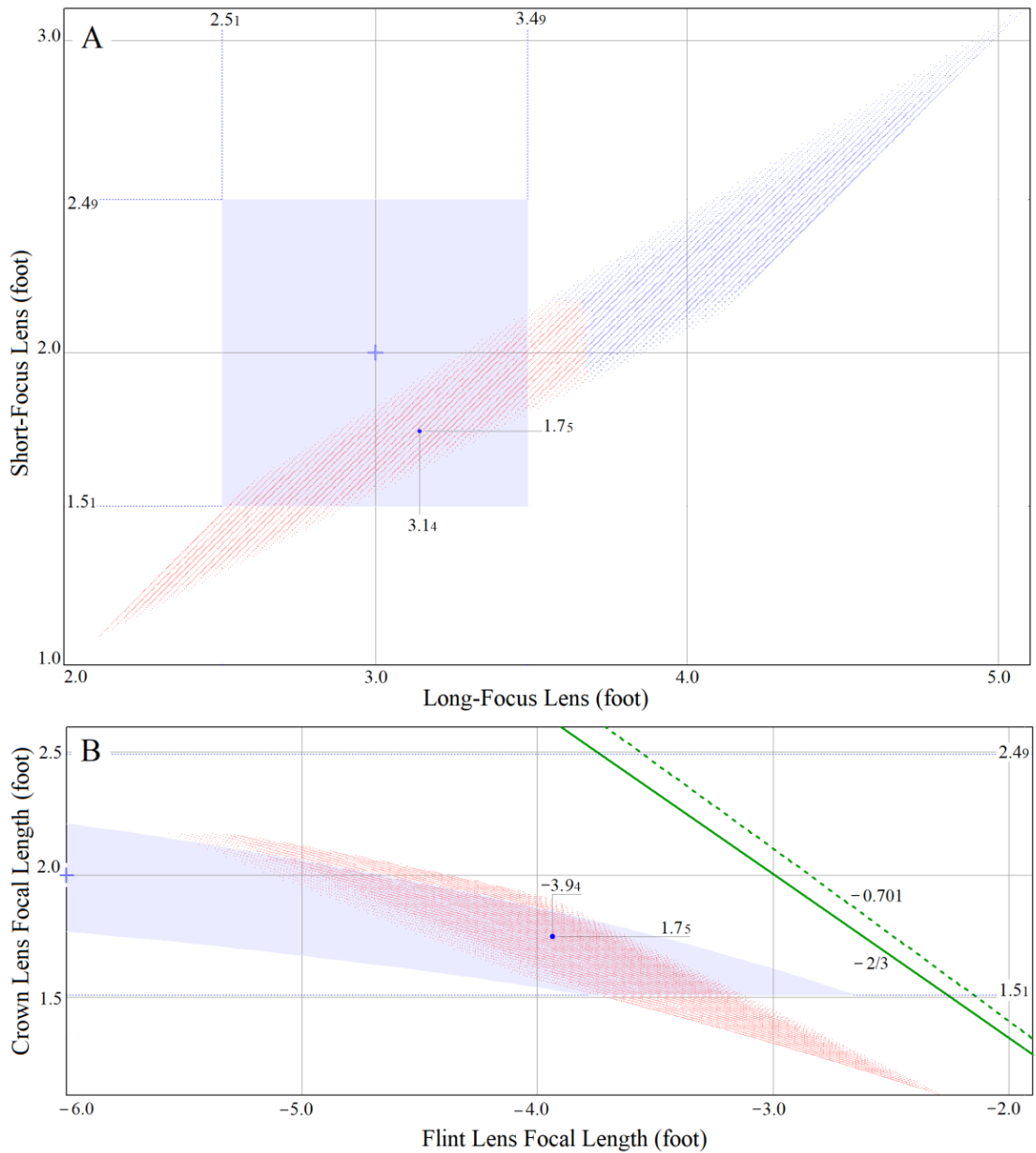

Figure I-2. A – Plot of short-focus vs long-focus lens; B – Focal length plot of crown vs flint lens. Each dot is one of variants of a telescope 'Adapted to Use at Sea' whose parameters satisfy the description. The bluish areas on the both plots correspond to *Quere* conditions for focal length of crown (short-focus) and compound (long-focus) lens. The single blue dot is a median point of intersection between the bluish area and a full set of red dots. The solid green line corresponds to the achromatic condition. The green dash line refers to achromatic condition for the glass pair used in the simulation (Appendix II).

---

[20] Average and standard deviation values of the full set of red dots are $(1.71 \pm 0.18)'$ for short and $(3.07 \pm 0.30)'$ for long focal length.

The bluish area on plot B is deformed due to non-linear dependence of the focal length for flint lens $F_f$ versus of focal lengths of short-focus (crown) $F_s$ and compound $F_l$ lenses: $F_f = -F_l F_s/(F_l - F_s)$.

The blue cross indicates the exact parameters currently recorded in *Quere* and is outside of the intersection area. This is an additional indication that these values were really rounded, and therefore, there is a systematic displacement of the blue cross.

Percentage of red dots, located above line $1.9_5$ for a short focal length in bluish area, is about 5%. This means 95% probability that Stedman′ and Champneys′ testimonies contained a numbers in form $'1.x'$ for a plano-convex crown lens, where $x$ means one of digits 5, 6, 7, 8 or 9, before they were replaced by $'2.'$

To left of the green line are located variants of compound lenses with chromatic undercorrection, and to right with overcorrection (Fig. I-2B). All possible variants of the telescope for sea (red dots) are distributed only on the left side and are not achromatic with significant undercorrection.

Based on this analysis, we can speak about an unambiguous accordance between a compound lens described in *Quere* and a two-component long-focus lens of Ayscough′ telescope for sea.

**Appendix II**

The comparison of different versions for long-focus $F_l \approx 957$ mm $(3.1_4')$ and short-focus $F_s \approx 533$ mm $(1.7_5')$ lenses with aperture 24mm was done. Herein, the glasses with next optical parameters for flint ($n_{ef}= 1.58228$, $\nu_{ef}= 42.12$)[21] and for crown ($n_{ec} = 1.52617$, $\nu_{ec} = 60.08$) were used. At simulation of all two-component lenses in Zemax software the distance between glasses $\delta = 3.2$ mm was added due to non-zero thickness of flip-mounts. Also, seven wavelengths with weighting coefficients corresponding to the photopic sensitivity of human eye were used.

The version naming is based on next rule: the first letter ′C′ or ′F′ means crown or flint glass, respectively. If the version name contains two letters, for example ′CF′, this means a crown-forward lens or flint-forward if the first letter is ′F′. Alphanumeric version numbers are shown in the bluish circles on Figures 3, II-1 and II-2.

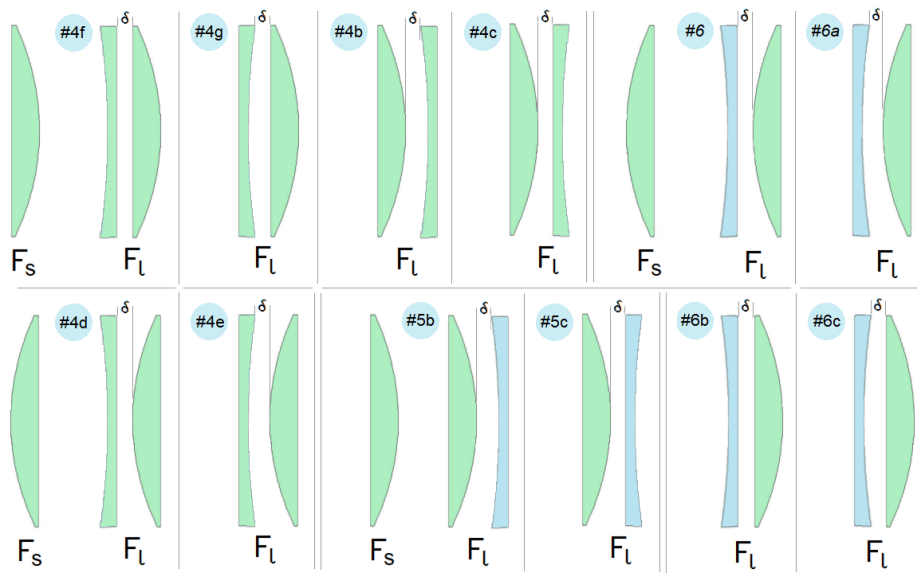

Figure II-1. Additional long-focus versions for design with flip out lenses. The remaining description is similar to Figure 3.

The simulation results were collected in Table II-1 for long-focus, and for short-focus lenses in Table II-2. For reference, a really achromatic lens is included in the Table II-1 with name ′CF_ach′, which is the best of all versions suitable for selected parameters and glass shapes. The versions with name C#1 mean a crown-singlet with the same focal length as the other versions listed in the same table. Versions #4 and #4a are comparable with C#1 in paraxial zone, but they are slightly worse due to larger field aberrations (basically coma). That is to say, with the chosen parameters, it is preferable to use a singlet lens instead of a two-component one. As expected, versions #5th and #6th with crown-flint pair have better chromatic compensation than the versions #4th and C#1 with one type of glass. A crown-flint compound lenses compatible with the description of telescope ′Adapted to Use at Sea′ have chromatism, which is compensated by only 1/3 of the value for singlet lens with the same focal length. Therefore, these versions are noticeably worse than a real achromat, especially near axis. Obviously, the versions with flint lens were more expensive than a singlet ones. Nevertheless, versions #5[th] and #6[th] could be released during the initial phase of project in 1753-55 years.

---

[21] The letter ′e′ at refractive index $n_e$ and Abbe number $\nu_e$ means e-line with wavelength 0.5461 μm.

Table II-1. Comparison of different versions of long-focus lenses[22]

| # | GEO (μm) | RMS (μm) | TSPH (μm) | TAXC (μm) | GEO (μm) | RMS (μm) | TSCO (μm) | TTCO (μm) | TAST (μm) | TLAC (μm) |
|---|---|---|---|---|---|---|---|---|---|---|
| | | | u = 0° | | | | u = 0.75° | | | |
| FC#6c | 73.5 | 22.8 | 21.6 | 124.6 | 97.5 | 24.7 | -7.1 | -21.2 | 1.9 | 1.7 |
| FC#6b | 75.0 | 22.9 | 24.4 | 124.7 | 93.6 | 24.0 | -5.3 | -15.9 | 1.9 | 1.8 |
| FC#6a | 71.5 | 22.7 | 17.7 | 124.4 | 84.9 | 23.1 | 2.9 | 8.8 | 2.1 | 1.5 |
| FC#6 | 73.0 | 22..8 | 20.5 | 124.5 | 91.8 | 23.7 | 4.7 | 14.2 | 2.1 | 1.6 |
| CF#5c | 84.7 | 23.9 | 42.7 | 125.3 | 110.4 | 25.9 | -7.2 | -21.5 | 2.0 | -0.6 |
| CF#5b | 79.9 | 23.4 | 33.4 | 125.2 | 100.4 | 24.6 | -5.5 | -16.4 | 2.1 | -0.5 |
| CF#5a | 67.2 | 22.8 | 8.4 | 125.5 | 78.9 | 23.0 | 2.8 | 8.3 | 2.1 | -0.9 |
| CF#5 | 62.7 | 22.8 | -0.9 | 125.3 | 79.3 | 23.4 | 4.5 | 13.4 | 2.1 | -0.8 |
| CF_ach | 30.1 | 10.7 | 52.3 | 7.3 | 57.4 | 16.2 | 8.3 | 25.0 | 2.1 | -3.0 |
| CC#4 | 94.7 | 34.3 | -1.9 | 188.5 | 111.0 | 34.6 | 4.6 | 13.8 | 2.1 | -0.5 |
| CC#4a | 99.0 | 34.2 | 8.5 | 188.6 | 110.5 | 34.3 | 2.7 | 8.1 | 2.1 | -0.6 |
| CC#4b | 111.2 | 34.5 | 32.5 | 188.4 | 131.1 | 35.2 | -5.3 | -16.0 | 2.1 | -0.2 |
| CC#4c | 116.5 | 34.9 | 42.9 | 188.5 | 142.3 | 36.2 | -7.2 | -21.7 | 2.0 | -0.3 |
| CC#4d | 105.1 | 34.3 | 20.3 | 188.7 | 124.1 | 34.8 | 4.8 | 14.5 | 2.1 | 1.6 |
| CC#4e | 103.4 | 34.2 | 17.2 | 188.4 | 116.4 | 34.4 | 2.8 | 8.5 | 2.1 | 1.5 |
| CC#4f | 107.2 | 34.4 | 24.2 | 189.0 | 125.5 | 35.1 | -5.2 | -15.6 | 1.9 | 1.9 |
| CC#4g | 105.5 | 34.3 | 21.1 | 188.8 | 129.8 | 35.5 | -7.2 | -21.6 | 1.9 | 1.8 |
| C#1 | 95.2 | 34.1 | 2.0 | 187.5 | 99.6 | 33.9 | 0.2 | 0.7 | 2.1 | 0.1 |

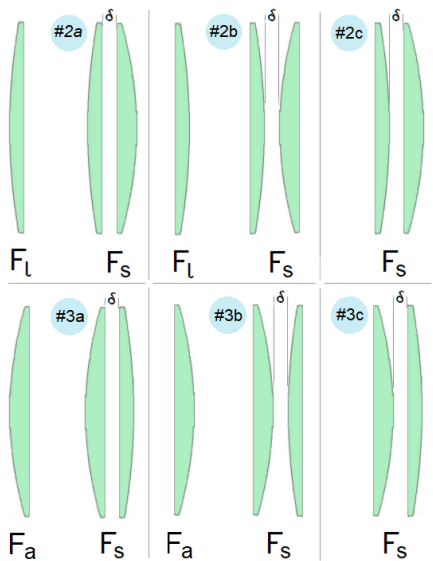

Figure II-2. Additional short-focus versions. The remaining description is similar to Figure 3.

Easy to see from Table II-2, as in the case of long-focus lens, the best short-focus versions CC#2 and CC#3 have no significant advantage versus of simple singlet lens with the same focal length. The decision to use a singlets instead of complex two-component lenses in later versions of the telescope for sea becomes more understandable, especially in a case of mass production.

In connection with the last statement, following question arises:
Could optician, for example, Eastland see residual chromatism of a long-focus compound lens at the chosen 16[x] magnification? At the moment, it is impossible to definitely answer this question, because the resolution of Eastland' eyes is unknown. It is only known that he was an optician with good experience and about 50 years old when was engaged by Ayscough. One of the best candidate of the same age whose eyes were examined and described is D. D. Maksutov. The resolution of his working eye are shown on Figure II-3. The best resolution is about 128 arc-seconds and is achieved with a pupil diameter of about 1.5 mm. Any imperfections in an image created by telescope lens are visible through the eyepiece with focal length $f_{ep}$ if their angular sizes are exceed the eye resolution $\delta_{ep}$:

$$\frac{250}{f_{ep}} \cdot \theta_{ch} \geq \delta_{ep} \qquad (II.1)$$

---

[22] Here in the table:
   GEO, RMS – envelope and RMS radii on the spot diagram, respectively. For reference, Airy radius is about 27 μm.
   TSPH, TAXC – spherical and axial chromatic aberration coefficients, respectively. Negative sign for the spherical and axial chromatic aberrations indicates about over-correction.
   TSCO, TTCO, TAST, TLAC – sagittal, tangential coma, astigmatism and lateral color aberration coefficients respectively at field angle u=0.75°. The first letter ′T′ in the coefficient names means transverse aberration.
   All aberration coefficients were calculated for e-line.

Here the eyepiece magnification $m_{ep} = 250/f_{ep}$, its focal length is expressed in millimeters; the angular size of chromatic aberration in arc-seconds $\theta_{ch} = 206265 \cdot TAXC/F$. Substituting into formula (II.1) expression for $\theta_{ch}$ and multiplying both parts of the inequality by $F$, we finally obtain formula for the minimum telescope magnification for chromatism observation:

$$M_{ch} \geq \frac{F^2 \cdot \delta_{ep}}{250 \cdot 206265 \cdot TAXC} \qquad (II.2)$$

Table II-2. Comparison of different versions of short-focus lenses[23]

| # | GEO (μm) | RMS (μm) | TSPH (μm) | TAXC (μm) | GEO (μm) | RMS (μm) | TSCO (μm) | TTCO (μm) | TAST (μm) | TLAC (μm) |
|---|---|---|---|---|---|---|---|---|---|---|
| | | | u = 0° | | | | u = 0.75° | | | |
| CC#3c | 100.4 | 34.0 | 12.5 | 187.2 | 114.8 | 34.2 | -3.7 | -11.1 | 2.0 | 0.8 |
| CC#3b | 96.1 | 34.0 | 4.0 | 187.2 | 105.6 | 34.0 | -2.0 | -6.1 | 2.0 | 0.7 |
| CC#3a | 99.5 | 34.0 | 10.7 | 187.1 | 110.6 | 34.0 | -2.6 | -7.8 | 2.0 | 0.7 |
| CC#3 | 95.1 | 34.0 | 2.3 | 187.0 | 101.4 | 33.9 | -0.9 | -2.8 | 2.0 | 0.6 |
| CC#2 | 95.4 | 34.0 | 2.8 | 187.0 | 101.7 | 33.9 | -0.9 | -2.8 | 2.0 | 0.5 |
| CC#2a | 98.4 | 34.0 | 8.6 | 187.0 | 107.8 | 34.0 | -2.0 | -6.0 | 2.0 | 0.6 |
| CC#2b | 97.1 | 34.0 | 6.1 | 187.2 | 108.6 | 34.0 | -2.6 | -7.9 | 2.0 | 0.6 |
| CC#2c | 100.1 | 34.0 | 11.9 | 187.2 | 114.6 | 34.2 | -3.7 | -11.1 | 2.0 | 0.7 |
| C#1 | 97.5 | 34.0 | 6.4 | 187.4 | 102.4 | 33.9 | 0.4 | 1.3 | 2.1 | 0.1 |

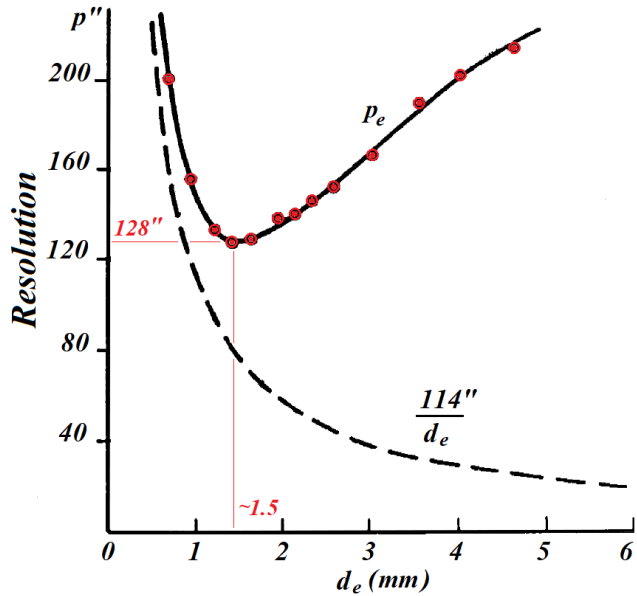

Figure II-3. Measured resolution of Maksutov' eye in arc-seconds (red dots) versus pupil diameter from his book (Fig. 58).[24]
The dash line is a theoretical resolution of diffraction-limited eye.

If we substitute in (II.2) the focal length 957 mm, the best resolution for eye 128″ and TAXC value from Table II-1 for any versions of crown-flint compound lens, then the minimum required magnification will be $18.1^x$. This means the chosen $16^x$ magnification really did not allow to observe the significantly undercorrected chromatism. To observe chromatism at this magnification, an angular resolution at least 113″ or better is required.

All above calculations are based on the geometrical optics approximation. Therefore, some results should be taken as an estimate, especially when the optical system is close to the diffraction limited. For the sake of completeness, the Peak-to-Valley (PTV) wave-front errors at different wavelengths for some versions of the long-focus lens are summarized in Table II-3. The calculated achromat (CF_ach) with chosen optical parameters and for given pair of crown-flint glasses is diffraction-limited for all used wavelengths. For this case, formula (II.2) for minimum magnification will give a preposterous result. As expected in terms of geometrical optics, the versions with flint glass #5[th] and #6[th] are better than the crown glass versions #4[th] and C#1. With only minor difference, the flint-forward (FC) versions are slightly better than the crown-forward (CF) ones. In contrast to the results in Table II-1 based on the geometrical optics, where the corresponding CF-versions are slightly better.

---
[23] All the same as in the description to Table II-1, only Airy radius is about 15 μm.
[24] D.D. Maksutov, 'Astronomical Optics', 2nd edition, (Nauka, Leningrad, 1979), p.159.

Table II-3. PTV for some versions of long-focus lens at different wavelength[25]

| # | u = 0º | | | | | | |
|---|---|---|---|---|---|---|---|
| | 0.4861 [0.177] | 0.5099 [0.501] | 0.5255 [0.801] | 0.5461 [0.984] | 0.5864 [0.800] | 0.6102 [0.500] | 0.6398 [0.177] |
| FC#6a | 0.8078 | 0.4272 | 0.2240 | 0.0255 | 0.3271 | 0.4693 | 0.6077 |
| FC#6 | 0.8089 | 0.4278 | 0.2243 | 0.0295 | 0.3275 | 0.4699 | 0.6084 |
| CF#5a | 0.8134 | 0.4301 | 0.2255 | 0.0120 | 0.3295 | 0.4727 | 0.6121 |
| CF#5 | 0.8119 | 0.4293 | 0.2250 | 0.0013 | 0.3289 | 0.4719 | 0.6110 |
| CF_ach | 0.1112 | 0.0944 | 0.0854 | 0.0755 | 0.0800 | 0.0818 | 0.0832 |
| CC#4 | 1.2214 | 0.6458 | 0.3386 | 0.0027 | 0.4949 | 0.7100 | 0.9193 |
| CC#4a | 1.2227 | 0.6466 | 0.3390 | 0.0123 | 0.4954 | 0.7107 | 0.9202 |
| C#1 | 1.2150 | 0.6425 | 0.3368 | 0.0029 | 0.4923 | 0.7063 | 0.9145 |

---

[25] Peak-to-Valley is calculated for the focal plane, where is minimum RMS of wave-front error. Wavelength is expressed in μm. The numbers in brackets [ ] indicate the photopic spectral sensitivity of human eye (CIE 2019) at the above wavelength. The green numbers are matched to condition of the diffraction-limited optics: wave-front error less than a quarter of wavelength.